\documentclass[sigconf,nonacm]{acmart}

\usepackage{array}
\usepackage{tabularx}
\usepackage{colortbl}
\usepackage{placeins}
\usepackage{xurl}
\newcommand{\ModelPrompts}{188}
\newcommand{\ModelCells}{1128}

\begin{document}

\title{Ask Again: Answer-List Instability in Generative Engines}

\author{
  \begin{tabular}{ccc}
    Björn~Busch-Geertsema &
    David~Antony~Selby    &
    Sebastian~Vollmer
  \end{tabular}
}
\affiliation{%
  \institution{University of Kaiserslautern--Landau (RPTU), \\
    German Research Center for Artificial Intelligence (DFKI)}
  \city{Kaiserslautern}
  \country{Germany}
}
\renewcommand{\shortauthors}{Busch-Geertsema et al.}

\begin{abstract}
  Visibility in generative search engines is measured by which entities appear
  in responses, and in what order. Ask an engine the same question twice and
  the answer list can change: a different ordering, or different items
  altogether. Across 60,000 responses, we find that answers to the same prompt
  name a different first item 32\% of the time, and even two answers with the
  same items disagree on which comes first. The first position is a
  measurement target of its own. A ``clear leader'' emerges after seven or
  eight runs, which are the budgets recommended for brand and source
  monitoring, but a ``close leader'' only about half the time. After 25 runs,
  a close leader is identified only 60\% of the time. Variability between
  engines is larger than within the same engine. For open-ended prompts, the
  gap is more than double that for determinate ones. Search queries and cited
  sites observed leave most of the disagreement unexplained. Visibility
  measurement should report, for each engine, which items appear and how often
  each comes first, each with its own quantification of uncertainty.
\end{abstract}

\ccsdesc[500]{Information systems~Web search engines}
\ccsdesc[500]{Information systems~Evaluation of retrieval results}
\ccsdesc[300]{General and reference~Measurement}

\keywords{generative search, generative engine optimization,
  answer stability, ranked lists, measurement reliability}

\maketitle

\begingroup
\begin{figure}
  \centering
  \includegraphics[width=\linewidth]{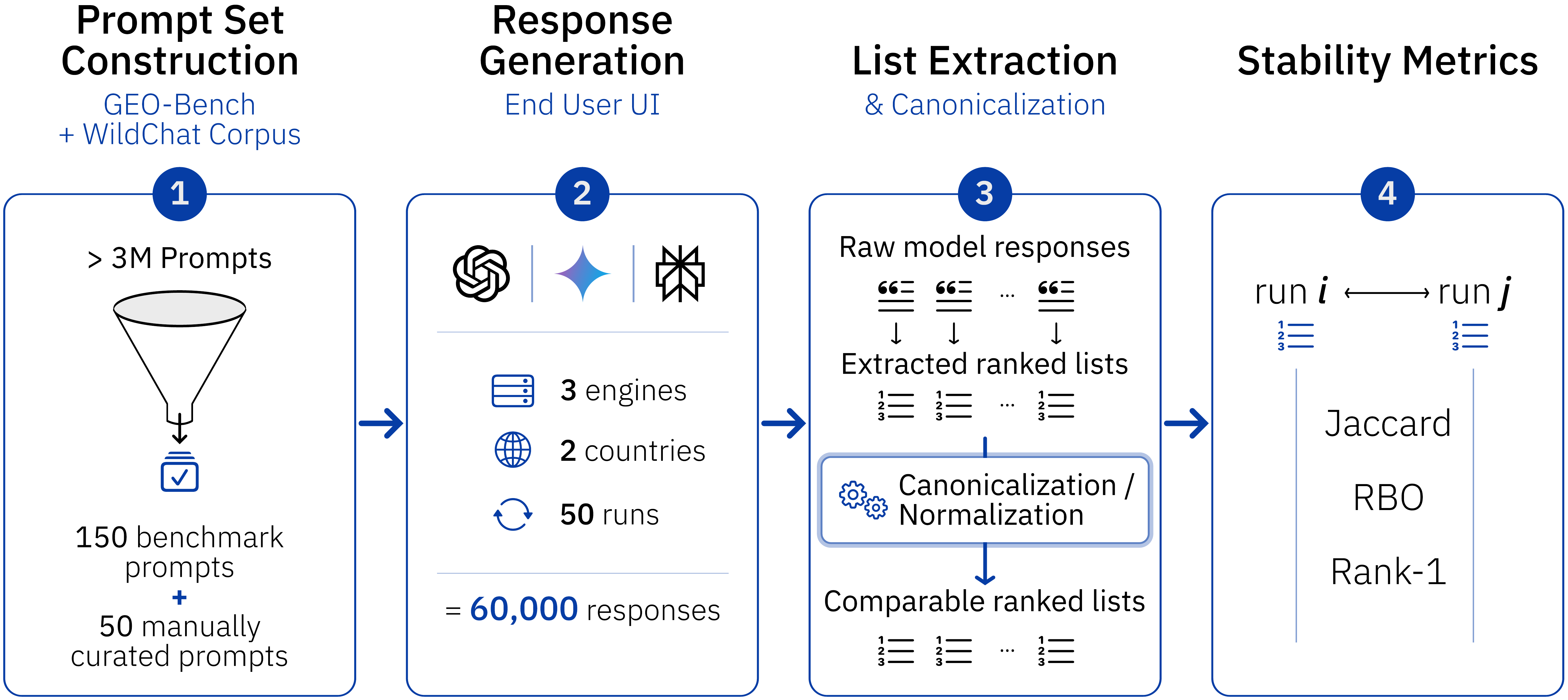}
  \caption{Experimental workflow. Each prompt is queried repeatedly and responses normalized.}
  \Description{Four-step study workflow. Prompt-set construction combines GEO-Bench and WildChat
    prompts to form 200 curated prompts. Response generation submits each prompt to ChatGPT, Gemini
    and Perplexity for Germany and the United States with 50 repetitions, yielding 60,000 responses. List extraction and canonicalization turn raw responses into comparable ranked answer lists. Stability metrics calculate Jaccard overlap, Rank-Biased Overlap, and rank-one agreement between repeated lists.}
  \label{fig:experiment-flow}
\end{figure}
\endgroup

\section{Introduction}

Generative engines answer information needs with composed responses informed
by web retrieval. These answers contain rankings: the first product,
institution, treatment, warning sign, or recommendation is usually more salient
than those that follow. In generative engine optimization (GEO), creators, brands, and researchers
measure whether an entity appears and in what order~\citep{aggarwal2024geo,
  puerto2025cseo,kim2026sageo,bagga2025egeo}. Such a measurement is meaningful
only if run-to-run variation is known. That variation is documented for cited
domains and brand mentions~\citep{sielinski2026uncertainty,
  schulte2026dontmeasureonce}, but not for what a user actually sees: an
answer that lists particular items in a particular order. Replication can
fail there in two ways. An engine can keep returning nearly the same list while
rotating the first item, or it can resample most of the list with each request.

We define answer-list stability by the variation among repeated anonymous
requests for one prompt, engine and country, treating the deployed
system as a whole: retrieval, generation, routing, experimentation, and
collection channel.

Our contributions:
\begin{itemize}
  \item We establish the first position as its own measurement target. An
        answer list can keep its items while the first item changes. When two items
        compete closely for first place, eight runs find the more frequent one only
        about half the time.
  \item Switching engine changes the answer more than rerunning one, most of
        all on open-ended prompts. One engine can consistently put first an item that
        another never lists.
  \item Stability depends far more on the individual prompt than on its topic
        or the country.
  \item A corpus of 60,000 responses from three deployed engines, with the
        extracted answer lists.
\end{itemize}

\section{Related Work}

GEO research has examined how modifying source content changes its visibility in
generated answers \citep{aggarwal2024geo,puerto2025cseo,wu2026autogeo}. Recent audits
establish that a single generative query is insufficient for measurement, with some
studies recommending repetition budgets of seven to eight runs for brand and source
monitoring \citep{schulte2026dontmeasureonce}, or context-dependent dynamic budgets
\citep{sielinski2026stable}. These budgets concern whether a brand or source appears,
or how cited domains rank by their share of citations, not the order of items within
one answer. Their data come from static documents and model APIs
\citep{aggarwal2024geo,puerto2025cseo,wu2026autogeo}, or from deployed engines sampled
through search platforms \citep{chen2025geoanalysis,kumar2026geoscale,sielinski2026uncertainty}
and web scrapers \citep{schulte2026dontmeasureonce}. Because we collect 50 runs per
cell, each within minutes, we can test whether these budgets suffice for what a user
actually sees: which items an answer lists and in what order.

Our work also relates to three further lines of research. First, while extensive
work audits the reliability, origin, and attribution of visible citations
\citep{liu2023verifiability,venkit2024searchengines,yang2025newssources,allaham2026synthetic},
engines may decouple these citations from the absorbed answer text \citep{zhang2026absorption}.
Second, whereas traditional search engine results pages exhibit volatility due to time,
personalization, location, and infrastructure
\citep{barilan2002performance,jimmy2018volatility,hannak2013personalization,klimansilver2015location,urman2022chance},
generative engines add retrieval and generation on top of this. Finally, at the generation layer,
language models are known to alter responses based on semantically equivalent inputs
\citep{novikova2025consistency,zhu2024promptbench} or reordered retrieved documents
\citep{zhang2026stablerag}. We measure the compound effect of these retrieval and generation instabilities on the final ranked list.

\section{Study Design}\label{sec:design}

We measure how much two runs of the same request agree. A \emph{cell} is
one prompt on one engine for one country, run 50 times. We sent each
prompt to three engines for two countries (Table~\ref{tab:glance},
Figure~\ref{fig:experiment-flow}). Within each cell we compare every pair
of responses and take the mean. We then average over cells and treat the
prompt as the unit of analysis.

\begin{table}[t]
  \caption{Study overview, further details in appendices~\ref{app:selection} (prompt selection), \ref{app:collection} (collection settings) and \ref{app:answer-lists} (outcomes)}
  \label{tab:glance}
  \begin{tabularx}{\columnwidth}{@{}>{\raggedright\arraybackslash}Xr@{}}
    \toprule
    \multicolumn{2}{@{}l}{\textit{Prompts}}                          \\
    GEO-Bench and WildChat pools screened & 3,209,860                \\
    Manually reviewed candidates          & 225                      \\
    Source-derived prompts kept           & 150                      \\
    Researcher-authored supplement        & 50                       \\
    \textbf{Study prompts}                & \textbf{200}             \\
    \addlinespace
    \multicolumn{2}{@{}l}{\textit{Design}}                           \\
    Engines (ChatGPT, Gemini, Perplexity) & 3                        \\
    Countries (DE, US)                    & 2                        \\
    Repetitions per cell                  & 50                       \\
    Cells                                 & 1,200                    \\
    \textbf{Responses collected}          & \textbf{60,000}          \\
    \addlinespace
    \multicolumn{2}{@{}l}{\textit{Outcomes}}                         \\
    Instrument failures excluded          & 653                      \\
    Non-empty answer lists                & 58,586                   \\
    Clarification requests                & 592                      \\
    Refusals                              & 168                      \\
    Partial response                      & 1                        \\
    \textbf{Outcomes analyzed}            & \textbf{59,347 (98.9\%)} \\
    \bottomrule
  \end{tabularx}
\end{table}

We used single-turn questions in English or German that a generative engine
would plausibly answer with a list. Prompts came from
GEO-Bench~\citep{aggarwal2024geo} and WildChat~\citep{zhao2024wildchat}.
Neither dataset targets such questions. GEO-Bench mixes several task types,
and WildChat has no fixed task. As WildBench did for
WildChat~\citep{lin2025wildbench}, we filtered both datasets and reviewed the
remaining candidates by hand, leaving their text unchanged. We added 50
prompts of our own to fill gaps in topic and language. Each of the five
topics (facts, health, science, localized, products) holds 40 prompts
(Appendix~\ref{app:selection}). The English and German prompts are not
translations of each other. Some prompts are subjective or broad
(Table~\ref{tab:prompt-examples}).

We collected responses from the public web interfaces of ChatGPT, Gemini and
Perplexity, without logging in. Requests went through Bright Data, a
commercial web-data platform, with the location set to Germany or the United
States. All responses were collected between 11 and 14 July
2026 (Appendix~\ref{app:collection}).

An LLM (gpt-4.1-mini) extracted the listed items from each response, in
order. Synonyms for the same item were merged across all responses to a prompt. 
If an item appeared more than once, occurrences after the first are
ignored. Every response from an engine is considered an \emph{outcome}, including clarification
requests and refusals, which we treat as \emph{no-answer outcomes}. Failed
requests are excluded (Table~\ref{tab:glance}, Appendix~\ref{app:answer-lists}).

\begin{table}[t]
  \caption{Stability measures: $S$ and $T$ are the two answer lists in a
    pair. Two empty lists score 1.}
  \label{tab:measures}
  \centering
  \small
  \begin{tabular}{@{}>{\raggedright\arraybackslash}p{0.29\columnwidth}>{\raggedright\arraybackslash}p{0.65\columnwidth}@{}}
    \toprule
    Measure             & Definition                                                \\
    \midrule
    Jaccard index $J$   & Share of items the two lists have in common,
    $|S\cap T|/|S\cup T|$, ignoring order.                                          \\
    \addlinespace
    Rank-biased overlap & Agreement in order, weighted towards the top of the list.
    We use extrapolated RBO with $p=.7$~\citep{webber2010rbo,corsi2024ties}.         \\
    \addlinespace
    Rank-one agreement  & $1$ when both lists name the same item first, $0$
    otherwise.                                                                      \\
    \bottomrule
  \end{tabular}
\end{table}

We score each pair of outcomes in a cell with the three measures in
Table~\ref{tab:measures} (Appendix~\ref{app:answer-lists}).

All intervals are 95\% bootstrap intervals over prompts, from 2,000
replicates stratified by topic, source and language unless stated otherwise.

Four LLMs from three developers coded the prompt features from the prompt text
alone, three times each. A feature enters the model only if the judges
agree, with Fleiss' $\kappa\geq.67$, on at least 20 prompts to which it
applies. On this basis, features included were subjectivity, explicit and requested list size,
geographic dependence and temporal sensitivity; specificity,
boundedness and ambiguity were excluded. We then fit a hierarchical Bayesian
ordered beta regression~\citep{kubinec2023orderedbeta} with one random intercept per prompt.
Covariates of the model are engine, topic, country, language, prompt source, log prompt length and the above-mentioned features extracted from the prompt text. We report average marginal
effects on the RBO scale (Appendix~\ref{app:measurement}).

\emph{Engine penalty} compares two engines $A$ and $B$ on the same
prompt and country,
\[
  P_{AB}=\tfrac{1}{2}(\bar s_{AA}+\bar s_{BB})-\bar s_{AB},
\]
where $\bar s$ is mean pairwise similarity within or between engines. A
penalty of zero means the two engines
are as interchangeable as two reruns of either one. The
\emph{locale penalty} does the same for Germany and the United States on one
engine. We
group engine comparisons into quarters by how \emph{determinate} the prompt
is, according to the within-engine RBO of the third engine. We call prompts in
the least determinate quarter \emph{open-ended}.

We also recorded the visible sources and, for ChatGPT and Perplexity, the
web-search queries the engines reported. For sources, we measure
how many source domains two outcomes share (Jaccard). ChatGPT and Gemini are scored
on cited sources and Perplexity on all sources it shows. We compare these
signals with answer stability within cells.

The \emph{leader} of a cell is the item its engine most frequently ranks first. 
We tested the run budgets recommended in prior
work~\citep{schulte2026dontmeasureonce} and larger budgets for two targets. For
RBO, we compared subsamples of each cell's outcomes with the value from all
outcomes. For the first item, we drew samples of $n$ runs from each cell's
outcomes and recorded how often they find the leader. We grouped cells by
the gap in percentage points between their two most frequent first items.
We call a leader \emph{clear} when this gap is $\geq 40$ points and
\emph{close} when it is $\leq 10$ (Appendix~\ref{app:sampling}).

\section{Results}

Two runs of the same request often return different answer lists and 
a different first item. Within a cell, two outcomes name a different first
item 32\% of the time and in most cells the first item changes at
least once (Table~\ref{tab:engine}). ChatGPT has a
higher RBO than Gemini and Perplexity, which differ little from each other
(Appendix~\ref{app:within}).

\begin{table}[t]
  \caption{Within-engine stability. `Items' is the mean/median number of
    distinct items in a non-empty list. `Varying' is the percentage of cells whose
    outcomes do not all share the first item. Brackets are 95\% intervals.}
  \label{tab:engine}
  \centering
  \setlength{\tabcolsep}{3pt}
  \newcommand{\ci}[1]{{\scriptsize[#1]}}
  \begin{tabular}{@{}lccccr@{}}
    \toprule
    Engine     & Items   & RBO          & Jaccard      & Rank-one     & Varying       \\
    \midrule
    ChatGPT    & 9.5/7   & .73          & .63          & .72          & 67\% \\[-2pt]
               &         & \ci{.70,.75} & \ci{.60,.65} & \ci{.68,.75} &               \\
    Gemini     & 8.3/6   & .68          & .60          & .66          & 74\% \\[-2pt]
               &         & \ci{.65,.70} & \ci{.57,.62} & \ci{.63,.70} &               \\
    Perplexity & 8.0/6   & .67          & .57          & .66          & 81\% \\[-2pt]
               &         & \ci{.64,.70} & \ci{.54,.60} & \ci{.63,.70} &               \\
    \midrule
    All        & 8.6/6   & .69          & .60          & .68          & 74\% \\[-2pt]
               &         & \ci{.67,.71} & \ci{.57,.62} & \ci{.65,.71} &               \\
    \bottomrule
  \end{tabular}
\end{table}

\begin{figure}[t]
  \centering
  \includegraphics[width=\linewidth]{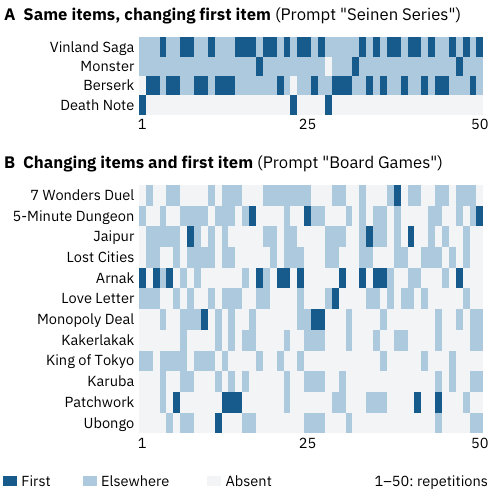}
  \caption{Two ways in which runs differ. Each panel shows one cell, and each
    column one outcome. A keeps its items while the first item changes, and B
    changes both. B shows the 12 most frequent of 105 items.}
  \Description{Two presence matrices across 50 repetitions. Nearly continuous
    rows in the seinen panel contrast with scattered presence in the board-game
    panel. Dark blue marks first position, light blue marks presence elsewhere,
    and pale grey marks absence. The board-game panel displays its 12 most
    frequent items. Arnak abbreviates Die verlorenen Ruinen von Arnak.}
  \label{fig:audited-shapes}
\end{figure}

The items in a list and its first item vary separately. About five times as
many cells name the same first item in every outcome as return the same list
every time. When two outcomes name different first items, both lists still
contain both items in about four of ten cases (Appendix~\ref{app:within}).
Even two outcomes with exactly the same items can name a different first
item (Figure~\ref{fig:audited-shapes}). This happens in seven of ten prompts
that have such pairs.

\begin{figure}[t]
  \centering
  \includegraphics[width=\linewidth]{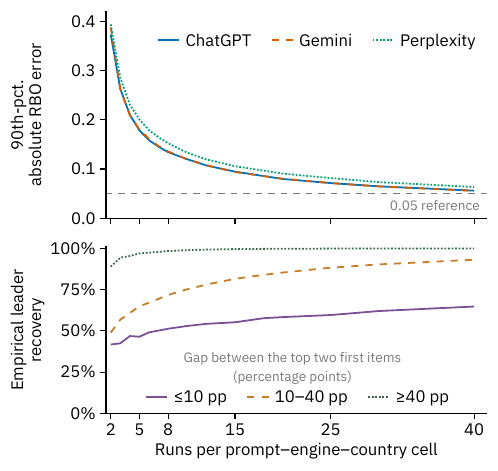}
  \caption{Sampling precision by target. Top: 90th-percentile absolute error
    of a cell's RBO. Bottom: how often a sample finds the cell's leader, grouped
    by the gap between the two most frequent first items.}
  \Description{Two line charts over 2 to 40 runs per cell. Top: the
    90th-percentile absolute RBO error falls from about .38 at two runs to about
    .06 at 40 runs. The ChatGPT and Gemini lines overlap, and Perplexity is
    slightly higher. Bottom: the share of samples that find the cell's leader, in
    three groups by the gap between the two most frequent first items. With a gap
    of at least 40 percentage points it is near 100\% from eight runs on. With a
    gap of 10 to 40 points it rises from about 49\% to 93\%. With a gap of at most
    10 points it rises from about 42\% to 65\%.}
  \label{fig:convergence}
\end{figure}

More runs steadily reduce the error of a cell's RBO
(Figure~\ref{fig:convergence}). Finding the leader is harder. Eight runs, as
recommended in prior work~\citep{schulte2026dontmeasureonce}, almost always
find a clear leader, and seven in ten cells have one. They find a close leader only about half the time, and even 25 runs find
one only about three times in five.

\begin{figure}[t]
  \centering
  \includegraphics[width=\linewidth]{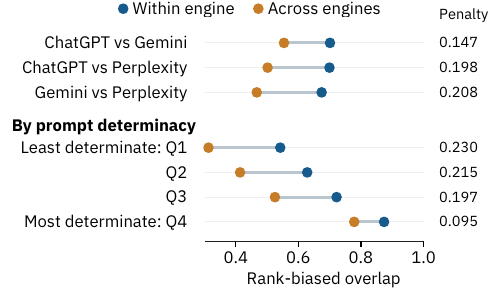}
  \caption{Switching engine costs more than rerunning one. Each row spans
    mean cross-engine RBO to the mean within-engine baseline, and the difference is
    the penalty. Grouping the same comparisons by determinacy (bottom group)
    shows larger penalties for less determinate prompts. Intervals, medians and
    per-country penalties are in Table~\ref{tab:engine-penalty-detail}.}
  \Description{A dot-and-span chart with seven rows on a rank-biased overlap
    axis. Each row has an amber dot at the cross-engine value and a blue dot at
    the higher within-engine value, joined by a grey bar whose length is the
    penalty. The three engine-pair rows span .554 to .701, .501 to .699, and .467
    to .674, giving penalties of .147, .198 and .208. The four prompt-determinacy
    rows run from least to most determinate, spanning .312 to .542, .414 to .628,
    .525 to .722 and .778 to .874, giving penalties of .230, .215, .197 and .095.
    The spans shorten and move right as prompts become more determinate.}
  \label{fig:engine-penalty}
\end{figure}

Switching engine changes the answer more than rerunning one
(Figure~\ref{fig:engine-penalty}). The engine penalty averages .18 RBO and is
positive in almost all (94\%) of engine comparisons. Comparisons involving Perplexity have
the largest penalties. The penalty also depends on how determinate the prompt
is. For open-ended prompts it is more than twice as large as for the most
determinate quarter. Switching country adds
much less on average. For about one prompt in six, however, the locale penalty
exceeds .10 on at least one engine (Table~\ref{tab:engine-penalty-detail}).

Engines can also disagree outright. For a prompt asking for three folk sports
from Karnataka, ChatGPT and Gemini name Kambala first in every outcome, while
Perplexity never lists it and names the same three other sports every time.
This is not an isolated case. In 14 prompts, one engine almost always puts
first an item that another engine never lists (Appendix~\ref{app:cases}).

Stability differs mainly from one prompt to the next. These differences account
for about two thirds of the variance in cell RBO,
while topic, engine and country together account for less than 5\%.
Everything we coded about a prompt, including its topic, language and
source, explains only about a fifth of the differences between prompts. Subjectivity is the only one with a clear
association. After adjustment, objective prompts have an RBO about .2 higher
than subjective prompts (Appendix~\ref{app:measurement}). Researcher-authored
prompts are also more stable than GEO-Bench and WildChat prompts.

The sources and search queries we can observe explain little of the
variation between runs. Within a cell, outcomes that share more source domains have only
slightly more similar answers, on every engine. On ChatGPT and Perplexity,
outcomes that issued the same search query are more alike than outcomes with
different queries, but they still differ. Two ChatGPT outcomes with the same
query and the same cited domains reach a mean RBO of .82
(Appendix~\ref{app:retrieval}).

Drift during collection does not explain the disagreement. Answer similarity
does not depend on the time between two runs, within a cell or between the
cells that a penalty compares. Mixed model labels show no consistent effect,
and exact repeats of an answer are rare (Appendix~\ref{app:robustness}).

\section{Discussion}

A visibility measurement in generative search samples one answer from the
many that an engine gives to the same request. What the measurement means
depends on four choices: what it counts, how many runs it takes, which
engines it covers and which prompts it asks. We discuss each choice below,
and close with what this means for those who try to change visibility.

Whether an item appears and whether it comes first are separate targets,
and in our data they vary separately (Figure~\ref{fig:audited-shapes}).
Guidance on repeated measurement so far concerns appearance.
\citet{schulte2026dontmeasureonce} derived run budgets from the standard
error of brand detection rates, and
\citet{sielinski2026uncertainty,sielinski2026stable} ranked domains by how
often they are cited across queries. Neither tracks which item an answer
puts first. A report should give, for each prompt and engine, how often an
item appears and how often it comes first, each with its own uncertainty.

How many runs a measurement needs depends on its target. More runs steadily
reduce the error of a cell's RBO, but whether a sample finds the leader
depends mainly on how close the two most frequent first items are
(Figure~\ref{fig:convergence}). The budgets of
\citet{schulte2026dontmeasureonce} were set for appearance. For the first
item they suffice when one item clearly leads, as it does in most cells,
but not when two items are close. This agrees with
\citet{sielinski2026stable} that no fixed budget fits every case, and it
points to one quantity that sets the budget: the gap between the two
leading items. Where that gap is small, a report should give how often each
item comes first rather than name a single leader.

Which engines a measurement covers is a separate choice from how often it
samples each one. Engines differ in which items they list and put first
(Figure~\ref{fig:engine-penalty}, Appendix~\ref{app:cases}).
\citet{chen2025geoanalysis} found that engines differ in the sources they
cite and argued for engine-specific strategies. Our results extend this to
the items an answer lists and puts first. A result from one engine does not
reliably carry over to another, so a report should name its engines and give results
for each. The difference between engines is largest for prompts whose
answers already vary within
an engine, and these prompts need coverage across engines most. Switching
country changes the answer less on average, but for some prompts it changes
it a lot, so it should be checked rather than assumed
(Table~\ref{tab:engine-penalty-detail}).

Stability depends mostly on the individual prompt. A prompt's topic or
wording tells little about how stable its answers will be
(Appendix~\ref{app:measurement}), so an average over a prompt set depends on
which prompts the set contains. Researcher-authored prompts are more stable
than GEO-Bench and WildChat prompts, so a prompt set written by researchers
may make engines look more stable than the questions people ask. For
GEO experiments, a change in one prompt's answers after an intervention can
be ordinary run-to-run variation, so a test needs many prompts, each
measured before and after.

For those who try to change visibility, being cited is not the same as being
listed. GEO-Bench measured visibility through the sentences of an answer
that cite a source~\citep{aggarwal2024geo}. In our data, outcomes that share
their sources, and even their search query, still give different answer
lists (Appendix~\ref{app:retrieval}). This fits the finding of
\citet{zhang2026absorption} that whether a page is cited and how much it
shapes the answer diverge. A gain in citations does not show that an item
moved up the answer list, so a study of optimization should measure the
list itself.

\section{Limitations}

We measured how often repeated requests agree, not whether the answers are
right. A stable answer can be wrong, and an unstable one can list several
valid options. Our results hold for anonymous requests to three web
interfaces, sent through one collection provider within one week of July
2026. We did not test logged-in, personalized, conversational or API use, and
we make no claims about change over days or weeks.

Because we measure the deployed system as a whole, we cannot tell whether
the variation comes from retrieval, generation, routing, experiments on the
provider's side or the collection channel. The sources and search queries we
observe are incomplete traces and do not settle this.

The prompt set is balanced by design, with equal numbers per topic and fixed
shares per language and source (Table~\ref{tab:prompt-grid}). Real use has no
such balance, so averages over prompts hold for this set only.

An LLM extracted the answer lists and merged different names for the same
item. Two annotators checked the lists of 180 runs against the raw answers
(Appendix~\ref{app:extraction-check}). They found no errors for first items
and errors that affect the full ordered list in 1.1\% of the inspected
runs. In the unchecked runs, missed or split items would make runs look less alike, and wrongly merged
items would make them look more alike.

\section{Conclusion}

Which item a generative engine names first needs to be measured on its own.
We sent the same prompts repeatedly to ChatGPT, Gemini and Perplexity and
compared the answer lists. The first item often changes from one run to the
next, even when the list keeps the same items. Switching engine changes the
answer more than rerunning one, most of all for prompts whose answers already
vary within one engine.

A visibility measurement should report, for each prompt and engine, how often
an item appears and how often it comes first, each with its uncertainty. The
number of runs it needs depends on how close the two most frequent first items
are. The budgets of \citet{schulte2026dontmeasureonce} suffice when one item
clearly leads. When two items are close, the measurement should report how
often each comes first rather than name a single leader. It should also name
the engines it covers and report each one separately.

\section*{Data Availability}
The prompts, all collected responses, their extracted answer lists and the
code for every analysis and figure are available at \url{https://anonymous.4open.science/r/ask-again-2DE6/}.

\subsection*{AI use statement}

We used LLMs in our methods as described: gpt-4.1-mini classified candidate prompts,
extracted the answer lists from the engine responses and merged different
names of the same items. Four LLMs coded the prompt features. These steps,
their settings and how we checked them are described in
Section~\ref{sec:design} and Appendices~\ref{app:selection},
\ref{app:answer-lists} and~\ref{app:measurement}. Two annotators checked a sample
of the extracted lists against the responses (Appendix~\ref{app:extraction-check}).

We used an AI coding assistant (Claude Code) to write analysis and
figure code and to explore the data. The authors set the initial direction, chose the analyses and directed the work. 
We reviewed the generated scripts and spot-checked their results. The released code
regenerates every number and figure in the paper (see Data Availability). 
We also used Claude Code for literature discovery, writing assistance and
review passes. The authors discussed and made all final decisions on the analyses, the
data exploration and the text of the paper. We did not use generative AI to create synthetic data.
We take responsibility for the final content of this work, including text, claims, or artifacts 
produced with the aid of generative AI.
\bibliographystyle{ACM-Reference-Format}
\bibliography{geo_stability}

\appendix

\section{Prompt selection}\label{app:selection}
We screened GEO-Bench and WildChat-4.8M in the stages of
Table~\ref{tab:flow-full}. In the manual review, we rejected, for example, prompts with a
single answer, an unbounded or time-bound answer, or no standalone
information need, such as code tasks and fragments
(Table~\ref{tab:prompt-examples}). We accepted 168 of the 225 candidates and
kept ten per source, language and topic. Two strata had fewer than ten
accepted prompts. For each we reviewed five replacements and used four in
total. The researcher-authored prompts fill five slots per language and topic
(Table~\ref{tab:prompt-grid}).

\begin{table}[ht]
  \caption{Prompt selection stages.}
  \label{tab:flow-full}
  \centering
  \footnotesize
  \setlength{\tabcolsep}{3pt}
  \begin{tabular}{@{}>{\raggedright\arraybackslash}p{0.21\columnwidth}>{\raggedright\arraybackslash}p{0.56\columnwidth}r@{}}
    \toprule
    Stage                      & Rule                                                                                      & Prompts      \\
    \midrule
    Input                      & GEO-Bench 10,000, WildChat-4.8M 3,199,860                                                 & 3,209,860    \\
    Language                   & English or German, from dataset metadata                                                  & 1,715,125    \\
    Length                     & 4 to 100 words                                                                            & 715,658      \\
    Intent                     & Information-seeking, by pattern rules                                                     & 497,573      \\
    Exact duplicates           & Removed                                                                                   & 326,483      \\
    Near duplicates            & Cosine similarity above .92 between all-MiniLM-L6-v2 embeddings,
    within source and language & 251,552                                                                                                  \\
    Presample                  & Stratified by source and language, seed 20260422                                          & 50,000       \\
    Classification             & gpt-4.1-mini, temperature 0, 3 failures                                                   & 49,997       \\
    Screen                     & Informational, in one of the five topics, stable over time, a list or a verifiable answer & 9,701        \\
    List shape                 & Classified as asking for a list                                                           & 3,474        \\
    Patterns                   & No code, translation or time-bound words                                                  & 3,260        \\
    Review                     & 15 per source, language and topic                                                         & 225          \\
    \rowcolor{black!10}
    Kept                       & Ten accepted per source, language and topic                                               & 150          \\
    \rowcolor{black!10}
    Researcher-authored        & Five per language and topic                                                               & 50           \\
    \midrule
    \textbf{Study prompts}     &                                                                                           & \textbf{200} \\
    \bottomrule
  \end{tabular}
\end{table}

\begin{table}[ht]
  \caption{Prompts by source and language. Each topic holds one fifth of every
    source and language.}
  \label{tab:prompt-grid}
  \centering
  \small
  \begin{tabular}{lrrr}
    \toprule
    Source              & English      & German      & Total        \\
    \midrule
    GEO-Bench           & 50           & 0           & 50           \\
    WildChat            & 50           & 50          & 100          \\
    Researcher-authored & 25           & 25          & 50           \\
    \midrule
    \textbf{Total}      & \textbf{125} & \textbf{75} & \textbf{200} \\
    \bottomrule
  \end{tabular}
\end{table}

\begin{table}[ht]
  \caption{Prompts rejected and accepted in the manual review.}
  \label{tab:prompt-examples}
  \centering
  \small
  \begin{tabular}{@{}>{\raggedright\arraybackslash}p{0.27\columnwidth}>{\raggedright\arraybackslash}p{0.67\columnwidth}@{}}
    \toprule
    Note          & Prompt                                                                              \\
    \midrule
    \multicolumn{2}{@{}l}{\textit{Rejected}}                                                            \\
    Single answer & ``who carried the us flag in the 2014 olympics''                                    \\
    Unbounded     & ``list of all known minerals''                                                      \\
    \addlinespace
    \multicolumn{2}{@{}l}{\textit{Accepted}}                                                            \\
    Ambiguous     & ``what are the main crops grown in the united states''                              \\
    Subjective    & ``what are some quirky sites from the early web''                                   \\
    Broad         & ``Was sind typische Wochenend-Unternehmungen eines Stadtbewohners in Deutschland?'' \\
    Decision task & ``Name five robot vacuum cleaners that work well in apartments with pets.''         \\
    \bottomrule
  \end{tabular}
\end{table}

\section{Collection}\label{app:collection}
\begin{table}[ht]
  \caption{Collection settings.}
  \label{tab:collection}
  \centering
  \footnotesize
  \begin{tabular}{@{}>{\raggedright\arraybackslash}p{0.34\columnwidth}>{\raggedright\arraybackslash}p{0.60\columnwidth}@{}}
    \toprule
    Setting            & Value                                             \\
    \midrule
    Provider           & Bright Data, one web scraper per engine           \\
    Engine pages       & chatgpt.com, gemini.google.com/app, perplexity.ai \\
    Country field      & DE or US                                          \\
    Prompt text        & Sent as selected, no added instructions           \\
    Runs per cell      & 50                                                \\
    Collection batches & 3, with 10, 40 and 150 prompts                    \\
    Active requests    & At most 50 provider snapshots                     \\
    Concurrent polls   & 20                                                \\
    Request timeout    & 180\,s                                            \\
    Response retries   & At most 2                                         \\
    First response     & 11 July 2026, 09:28 UTC                           \\
    Last response      & 14 July 2026, 07:45 UTC                           \\
    \bottomrule
  \end{tabular}
\end{table}

For ChatGPT, the scraper's \texttt{web\_search} option was set to
\texttt{false}, so web search was not forced and ChatGPT decided on its own
whether to search. It searched in most outcomes (Table~\ref{tab:retrieval}). The interfaces expose no sampling
temperature. Cell batches ran concurrently. Engine and country batches were
neither synchronized nor randomly ordered.

\section{Answer lists and scoring}\label{app:answer-lists}
\begin{table}[ht]
  \caption{Extraction and canonicalization settings.}
  \label{tab:extraction}
  \centering
  \footnotesize
  \begin{tabular}{@{}>{\raggedright\arraybackslash}p{0.30\columnwidth}>{\raggedright\arraybackslash}p{0.64\columnwidth}@{}}
    \toprule
    Setting            & Value                                                                         \\
    \midrule
    \multicolumn{2}{@{}l}{\textit{Extraction}}                                                         \\
    Model              & gpt-4.1-mini, OpenAI Batch API, temperature 0                                 \\
    Input              & Prompt, prompt language and response text                                     \\
    Output             & Item labels in text order and an outcome class, as structured output          \\
    Failed extractions & Resubmitted once for 150 of the 200 prompts                                   \\
    \addlinespace
    \multicolumn{2}{@{}l}{\textit{Canonicalization, per prompt}}                                       \\
    Step 1             & Merge labels that match after case folding                                    \\
    Step 2             & gpt-4.1-mini, temperature 0, groups labels in chunks of 120, up to ten rounds \\
    Unmatched labels   & Kept as items of their own                                                    \\
    Manual merges      & None                                                                          \\
    Canonical items    & 15,927 over 200 prompts                                                       \\
    \bottomrule
  \end{tabular}
\end{table}

The extractor returned short labels, not quoted text spans
(Table~\ref{tab:extraction}). It also assigned every outcome class except
parser-uncertain (Table~\ref{tab:outcome-classes}). Four clarification
requests, all from Perplexity, also list items, and we analyze them with
these items.

\begin{table}[ht]
  \caption{Outcome classes and the number of excluded responses in each.}
  \label{tab:outcome-classes}
  \centering
  \footnotesize
  \setlength{\tabcolsep}{3pt}
  \begin{tabular}{@{}>{\raggedright\arraybackslash}p{0.25\columnwidth}>{\raggedright\arraybackslash}p{0.54\columnwidth}r@{}}
    \toprule
    Class                 & Definition                              & Excluded \\
    \midrule
    \multicolumn{3}{@{}l}{\textit{Analyzed}}                                   \\
    Answer list           & At least one item                       &          \\
    Clarification request & The engine asks a question back         &          \\
    Refusal               & The engine declines to answer           &          \\
    Partial response      & The response breaks off before any item &          \\
    \addlinespace
    \multicolumn{3}{@{}l}{\textit{Excluded}}                                   \\
    Parser-uncertain      & The extractor returned no usable result & 581      \\
    Missing answer        & The response holds no answer            & 48       \\
    Provider error        & The response is an error message        & 22       \\
    Login or consent page & The response is a login or consent page & 2        \\
    \bottomrule
  \end{tabular}
\end{table}

A pair of a no-answer outcome and a non-empty list scores 0 on every
measure. For rank-one agreement and the first-item analyses, a no-answer
outcome counts as a first item of its own. Across the 1,413,800 within-cell
pairs of non-empty lists, 82.9\% of the RBO weight falls on positions that
both lists reach, 4.7\% on positions between the ends of the shorter and the
longer list, and 12.4\% beyond the longer list, where RBO is extrapolated.

\subsection{List check}\label{app:extraction-check}
Two annotators checked the extracted lists of 180 analyzed outcomes against the
response text. The sample held three outcomes for each combination of engine,
country, prompt language and topic (seed 20261004). Annotators saw the prompt,
the response with the extracted items highlighted and the extracted list,
with the canonical label wherever it differed. Engine and country were
hidden. Each annotator marked every list, and separately its first item, as correct or not correct. A list marked
not correct needed at least one reason and a note. The reasons were a missing
item, an item that is not part of the answer, a wrong label, a wrong merge or
split, a wrong order, a wrong canonical label, and other. The annotators agreed
on 178 of the 180 list verdicts (98.9\%). Cohen's $\kappa$ is 0 because one annotator
marked every list as correct. Both annotators marked every first item as correct,
so $\kappa$ is undefined for the first item.

\section{Prompt features and model}\label{app:measurement}
The four judges were gpt-4.1-mini and gpt-5.2, called through the OpenAI API,
and two open-weight models run locally with Ollama (\texttt{qwen3.8:27b} and
\texttt{muse-glimmer}). Each judge coded every prompt three times, first at
temperature 0 and then twice at temperature 1. gpt-5.2 takes no temperature
setting. The rubric defines every level of each feature and gives
contrasting examples that are not study prompts. A judge's label is its most
frequent label over its three runs. A prompt's label is the majority over the
four judges. Ties go to the most frequent of the 12 run labels and stay
missing if still tied. Agreement is Fleiss' $\kappa$ over the judges' labels,
on the prompts that at least two judges coded (Table~\ref{tab:features}). With
a fifth judge, \texttt{nemotron-3.5-lightning}, subjectivity reaches
$\kappa=.609$.

\begin{table}[ht]
  \caption{Agreement among the four judges. A feature enters the model if
    $\kappa\geq.67$ on at least 20 prompts.}
  \label{tab:features}
  \centering
  \small
  \begin{tabular}{@{}lrrl@{}}
    \toprule
    Feature               & $\kappa$ & Prompts & In model \\
    \midrule
    Requested list size   & 1.00     & 22      & yes      \\
    Explicit list size    & .89      & 200     & yes      \\
    Geographic dependence & .80      & 200     & yes      \\
    Temporal sensitivity  & .72      & 200     & yes      \\
    Subjectivity          & .67      & 200     & yes      \\
    Specificity           & .50      & 200     & no       \\
    Boundedness           & .43      & 200     & no       \\
    Ambiguity             & .15      & 200     & no       \\
    \bottomrule
  \end{tabular}
\end{table}

Requested list size is 0 when the judges agreed that the prompt states no
size. ``Unclear'' is a level of its own. Six prompts have no subjectivity
label, five no temporal sensitivity label and one no explicit-size label. We
excluded these prompts from the model, which leaves \ModelPrompts{} prompts
and \ModelCells{} within-engine cells.

The model is an ordered beta regression with a logit link, a random
intercept per prompt and the covariates listed in Section~\ref{sec:design}.
Coefficient priors are $\mathcal{N}(0,1)$ on the link scale, the precision
has an exponential prior with rate .1, and the cutpoints a Dirichlet$(1,1,1)$
prior. We ran four chains of 8,000 iterations, half of them warmup, with
seed 20260715. The largest $\hat R$ is 1.006, the smallest bulk ESS is 1,581,
and no transition diverged. The central 95\% posterior-predictive ranges
cover the observed mean, standard deviation, 10th, 50th and 90th
percentiles, and the shares of scores at 0 and 1. Intercept and
random-effect priors and the full formula are saved with the fit.

\begin{table}[ht]
  \caption{Average marginal effects on within-engine RBO with 95\% posterior
    intervals (\ModelPrompts{} prompts). Each level is compared with the
    reference in parentheses. Objective minus subjective is $+.22$
    [$+.12,+.32$], and researcher-authored minus WildChat is $+.13$
    [$+.07,+.19$]. Engine, prompt length and requested list size are in the
    model but not shown. ``Unclear'' levels are not shown.}
  \label{tab:model}
  \centering
  \small
  \begin{tabular}{@{}lr@{}}
    \toprule
    Level (reference)                         & Difference in RBO     \\
    \midrule
    \textit{Topic} (facts)                    &                       \\
    \quad Health                              & $-.06$ [$-.15,+.03$]  \\
    \quad Localized                           & $-.06$ [$-.16,+.04$]  \\
    \quad Products                            & $-.07$ [$-.16,+.02$]  \\
    \quad Science                             & $-.04$ [$-.12,+.05$]  \\
    \textit{Subjectivity} (mixed)             &                       \\
    \quad Objective                           & $+.12$ [$+.06,+.18$]  \\
    \quad Subjective                          & $-.10$ [$-.20,-.003$] \\
    \textit{Source} (GEO-Bench)               &                       \\
    \quad Researcher-authored                 & $+.15$ [$+.06,+.24$]  \\
    \quad WildChat                            & $+.02$ [$-.07,+.10$]  \\
    \textit{Geographic dependence} (explicit) &                       \\
    \quad Implicit                            & $-.05$ [$-.18,+.08$]  \\
    \quad None                                & $-.06$ [$-.15,+.03$]  \\
    \textit{Temporal sensitivity} (current)   &                       \\
    \quad Explicit time frame                 & $+.08$ [$-.06,+.20$]  \\
    \quad Stable                              & $+.06$ [$-.02,+.13$]  \\
    \addlinespace
    List size stated (no)                     & $+.09$ [$-.01,+.18$]  \\
    English (German)                          & $+.02$ [$-.05,+.08$]  \\
    US (DE)                                   & $.00$ [$-.02,+.01$]   \\
    \bottomrule
  \end{tabular}
\end{table}

The coefficients for log prompt length and requested list size have 95\%
intervals that include zero. To compare prompts with other factors, we split
the total sum of squares of cell RBO by grouping. Prompts account for 69.0\%
(one-way intraclass correlation .63), topics for 3.0\%, engines for 1.2\%
and countries for less than 0.01\%. With 200 prompts of six cells each,
grouping by prompt reaches about 17\% by chance. A linear regression of
prompt-mean RBO on topic, language, source, log prompt length and the coded
features explains 27\% of the variance between the \ModelPrompts{} prompts
(adjusted $R^2=.19$).

\section{Sampling precision}\label{app:sampling}
For the RBO error (Table~\ref{tab:sampling}), we divided each error by $\sqrt{1-n/N}$, a first-order
finite-population correction. As a check at 25 runs, we split each of the
1,044 cells with 50 outcomes into two disjoint halves. After scaling, the
errors from these halves differ from the corrected errors by less than .002.

\begin{table}[ht]
  \caption{Sampling settings for the two targets in
    Figure~\ref{fig:convergence}.}
  \label{tab:sampling}
  \centering
  \footnotesize
  \setlength{\tabcolsep}{3pt}
  \begin{tabular}{@{}>{\raggedright\arraybackslash}p{0.22\columnwidth}>{\raggedright\arraybackslash}p{0.35\columnwidth}>{\raggedright\arraybackslash}p{0.37\columnwidth}@{}}
    \toprule
                             & RBO error                             & Leader                                     \\
    \midrule
    Cells                    & 1,188 with at least 41 outcomes       & 1,033 with 50 outcomes and a unique leader \\
    Runs $n$                 & 2--8, 10, 12, 15, 20, 25, 30, 40      & 2--6, 8, 10, 12, 15, 18, 20, 25, 30, 40    \\
    Samples per cell and $n$ & 200                                   & 10,000                                     \\
    Drawn                    & Without replacement                   & With replacement                           \\
    Reference                & Mean pairwise RBO of all $N$ outcomes & Empirical leader                           \\
    Seed                     & 20260715                              & 20260922                                   \\
    \bottomrule
  \end{tabular}
\end{table}

\begin{table}[ht]
  \caption{90th-percentile absolute RBO error after finite-population
    correction. Ranges span the three engines.}
  \label{tab:convergence}
  \centering
  \small
  \begin{tabular}{rr}
    \toprule
    Runs & Error range \\
    \midrule
    5    & .179--.200  \\
    8    & .134--.152  \\
    15   & .094--.106  \\
    25   & .071--.082  \\
    40   & .056--.064  \\
    \bottomrule
  \end{tabular}
\end{table}

For the leader, we keep the 64 cells with a no-answer outcome and leave out
the 11 cells in which the two most frequent first items tie. A sample finds the leader when its most frequent first item is the
leader. Ties within a sample go to the lexicographically greatest canonical
ID. We grouped cells by the gap between their two most frequent first items:
at most 10 percentage points (88 cells), more than 10 and less than 40 (214
cells), and at least 40 (731 cells). Each curve in the
bottom panel of Figure~\ref{fig:convergence} is the mean over the cells of a group.

These samples treat the 50-run frequencies as exact. As a check, we drew
10,000 sets of first-item frequencies per cell from a Dirichlet distribution
with the cell's counts plus one half (seed 20261004). We scored one sample
per draw against the item that leads in that draw, with ties broken at
random. This check includes the 11 tied cells. For gaps of at most 10
percentage points, 8 runs then find the leader 53\% of the time and 25 runs
65\% of the time, against 51\% and 60\% in the main analysis.

\section{Within-engine stability}\label{app:within}
Engine differences in within-engine RBO use paired bootstrap intervals that
resample prompts within topic, source and language, with both countries of a
prompt kept together (2,000 replicates, seed 20260914). ChatGPT minus Gemini
is $+.050$ [$+.031,+.071$], ChatGPT minus Perplexity is $+.053$
[$+.026,+.080$], and Gemini minus Perplexity is $+.003$ [$-.023,+.029$].

In 315 cells, every outcome names the same first item. In 65 cells, every
outcome returns the same list. Among the 277,711 within-cell pairs of
non-empty outcomes whose lists contain the same items, 21.5\% differ in order
and 8.1\% [5.7, 10.7] name a different first item. Such pairs occur in 193
prompts, and first-item disagreement occurs in 136 of them. Among the 444,609
within-cell pairs of non-empty outcomes that name different first items, both
lists contain both first items in 40.8\% [37.0, 44.7], one list contains the
other's in 36.3\%, and neither does in 22.9\%. Both intervals resample prompts
and use the ratio of summed pair counts.

\section{Engine and locale penalties}\label{app:penalties}
The engine penalty was positive in 1,133 of the 1,200 comparisons. Each
determinacy quartile holds 300 comparisons. The locale penalty exceeded .10
in 42 of the 600 prompt and engine combinations, from 34 prompts
(Table~\ref{tab:engine-penalty-detail}). When we shuffle the country labels
within each prompt and engine (200 shuffles), no combination exceeds .10.

\begin{table}[ht]
  \caption{Penalties in RBO. Self is the within-engine baseline (within-country
    in D), and Cross the similarity across engines (across countries in D).
    Intervals in A and C resample prompts. Intervals in B and D also resample
    runs within cells. C, G and P stand for ChatGPT, Gemini and Perplexity.
    Penalties use unrounded values.}
  \label{tab:engine-penalty-detail}
  \centering
  \footnotesize
  \setlength{\tabcolsep}{4pt}
  \begin{tabular}{lrrrr}
    \toprule
                              & Self & Cross & Penalty                 & Median \\
    \midrule
    \multicolumn{5}{l}{\textit{A. Engine penalty}}                              \\
    All comparisons           & .691 & .507  & .184 [.168,.201]        & .139   \\
    ChatGPT vs Gemini         & .701 & .554  & .147 [.130,.165]        & .103   \\
    ChatGPT vs Perplexity     & .699 & .501  & .198 [.177,.219]        & .153   \\
    Gemini vs Perplexity      & .674 & .467  & .208 [.184,.229]        & .171   \\
    \addlinespace
    Least determinate         & .542 & .312  & .230 [.204,.254]        & .201   \\
    Second quartile           & .628 & .414  & .215 [.191,.239]        & .166   \\
    Third quartile            & .722 & .525  & .197 [.172,.223]        & .165   \\
    Most determinate          & .874 & .778  & .095 [.070,.126]        & .029   \\
    \midrule
    \multicolumn{5}{l}{\textit{B. Engine penalty by engine pair and country}}   \\
    ChatGPT vs Gemini, DE     & .702 & .549  & .153 [.134,.174]        &        \\
    ChatGPT vs Gemini, US     & .700 & .559  & .141 [.123,.160]        &        \\
    ChatGPT vs Perplexity, DE & .698 & .500  & .198 [.177,.221]        &        \\
    ChatGPT vs Perplexity, US & .700 & .503  & .198 [.176,.220]        &        \\
    Gemini vs Perplexity, DE  & .677 & .463  & .214 [.191,.237]        &        \\
    Gemini vs Perplexity, US  & .671 & .470  & .201 [.179,.223]        &        \\
    \midrule
    \multicolumn{5}{l}{\textit{C. Differences between engine pairs}}            \\
    C--P minus C--G           &      &       & $+.051$ [$+.029,+.072$] &        \\
    G--P minus C--G           &      &       & $+.060$ [$+.040,+.080$] &        \\
    G--P minus C--P           &      &       & $+.010$ [$-.009,+.029$] &        \\
    \midrule
    \multicolumn{5}{l}{\textit{D. Locale penalty (DE vs US)}}                   \\
    ChatGPT                   & .726 & .698  & .028 [.019,.037]        & .003   \\
    Gemini                    & .676 & .637  & .039 [.029,.049]        & .013   \\
    Perplexity                & .673 & .667  & .006 [.003,.009]        & .000   \\
    \bottomrule
  \end{tabular}
\end{table}

\begin{table}[ht]
  \caption{Mean engine penalty under other choices.}
  \label{tab:penalty-checks}
  \centering
  \small
  \begin{tabular}{@{}lrr@{}}
    \toprule
    Variant                             & Comparisons & Penalty \\
    \midrule
    Main analysis (RBO)                 & 1,200       & .184    \\
    Without no-answer outcomes          & 1,198       & .180    \\
    Without researcher-authored prompts & 900         & .183    \\
    Lists cut to five items             & 1,030       & .190    \\
    Jaccard instead of RBO              & 1,200       & .174    \\
    Rank-one agreement instead of RBO   & 1,200       & .198    \\
    \bottomrule
  \end{tabular}
\end{table}

For lists cut to five items, we kept only lists with at least five items and
only comparisons in which both engines have at least two such lists
(Table~\ref{tab:penalty-checks}).

\subsection{Engine-specific leaders}\label{app:cases}
An engine-specific leader is an item that comes first in at least 80\% of a
cell's outcomes and appears in no outcome of another engine for the same
prompt and country. Both cells need at least 40 outcomes, and the other
engine's cell at least 40 non-empty ones. This holds for 24 engine
comparisons in 14 prompts.

For the Karnataka prompt, we checked all 300 raw answers against their
extracted lists. Searches for spelling variants of Kambala agree with its
extracted presence in every answer. In 194 of the 200 ChatGPT and Gemini
answers, the answer list starts with Kambala. In the other six, a source or
citation marker comes first. We inspected these by hand, and Kambala is the
first item in all six. All 100
Perplexity answers list Chinni Dandu, Bugari and Kabaddi. The six cells were
collected on 11 and 12 July in batches that were not synchronized.

\section{Sources and search queries}\label{app:retrieval}
Perplexity marks no source as cited. Domain overlap is the Jaccard index of
the source domains of two outcomes that both show a source, averaged over
cells with at least two such outcomes. For Gemini, 213 of 400 cells qualify.
Concentration is the Gini coefficient of domain counts over all outcomes of
an engine. The correlation relates domain overlap to answer RBO over pairs of
outcomes that both show a source, after subtracting each cell's mean from
both. Gemini reports no search queries.

\begin{table}[ht]
  \caption{Sources and search queries by engine. Shares, counts and overlap are
    means over cells. The query gap is the mean within-cell difference in RBO
    between pairs with the same query and pairs with different queries.}
  \label{tab:retrieval}
  \centering
  \footnotesize
  \setlength{\tabcolsep}{4pt}
  \begin{tabular}{@{}lrrr@{}}
    \toprule
                                     & ChatGPT & Gemini & Perplexity \\
    \midrule
    \multicolumn{4}{@{}l}{\textit{Sources}}                          \\
    Outcomes with a source           & 92.4\%  & 35.6\% & 67.3\%     \\
    Median sources per outcome       & 3.7     & 1.4    & 7.0        \\
    Domain overlap (Jaccard)         & .539    & .378   & .853       \\
    Domain concentration (Gini)      & .806    & .762   & .583       \\
    Correlation with answer RBO      & .110    & .060   & .096       \\
    \addlinespace
    \multicolumn{4}{@{}l}{\textit{Search queries}}                   \\
    Web search flag set              & 96.7\%  & --     & --         \\
    Outcomes with a query            & 97.5\%  & 0\%    & 67.3\%     \\
    Median distinct queries per cell & 12      & --     & 1          \\
    Query gap in RBO                 & .060    & --     & .120       \\
    \bottomrule
  \end{tabular}
\end{table}

ChatGPT used a single query in 2 of its 400 cells, and Perplexity in 298.
Pooled over cells, two ChatGPT outcomes with the same query and the same
cited domains have a mean RBO of .819 (40,484 pairs). With the same query
but different cited domains, the mean is .774 (72,601 pairs).

\section{Drift, model labels and repeated answers}\label{app:robustness}
The median time from the first to the last outcome of a cell is 6.7 minutes
for ChatGPT, 3.0 for Gemini and 3.2 for Perplexity. The 90th percentiles are
11.1, 6.3 and 6.0 minutes. Five of the 1,200 cells took more than an hour.
Within cells, we correlated the time between two outcomes with their RBO
after subtracting each cell's mean from both. The pooled correlations are
$-.002$ for ChatGPT, $+.008$ for Gemini and $+.009$ for Perplexity, each over
about 480,000 pairs. Cells that a penalty compares were collected a median
of 1.5 hours apart and at most 18 hours apart. The mean engine penalty is
.181 for cells collected less than an hour apart, .186 for one to six hours
and .188 for six hours or more. The mean locale penalty is .024, .023 and
.025 in the same bands.

The provider model label is \texttt{gpt-5-5} in 19,996 ChatGPT records and
\texttt{gpt-5-5-thinking} in four. Gemini records show \texttt{Flash} in
19,861, \texttt{3.5 Flash} in 93 and no label in 46. Perplexity records carry
no label. In 101 cells (97 Gemini, 4 ChatGPT) the label differs between
outcomes. In 93 of them, the minority labels cover at most two outcomes.
Pairs with different labels are less similar than pairs with the same label
on ChatGPT (RBO .929 against .953, 196 pairs) and more similar on Gemini
(.705 against .685, 6,530 pairs).

Exact repeats of the full answer text, ignoring case and whitespace, make up
at most 0.062\% of within-cell pairs (Gemini), against 0.046\% for ChatGPT
and 0.009\% for Perplexity. The largest group of identical answers in one
cell has 19 outcomes.

\end{document}